\documentclass{sapm}
\usepackage{graphicx}
\expandafter\let\csname equation*\endcsname\relax
\expandafter\let\csname endequation*\endcsname\relax
\usepackage{amsmath}
\usepackage{amssymb, cancel}

\newcommand{\beq}{\begin{equation}}
\newcommand{\eeq}{\end{equation}}

\newcommand{\ri}{r^{(\mathrm{in})}}
\newcommand{\ro}{r^{(\mathrm{out})}}
\newcommand{\si}{s^{(\mathrm{in})}}
\newcommand{\so}{s^{(\mathrm{out})}}
\newcommand{\rbar}{r^{(0)}}
\newcommand{\rtil}{r^{(1)}}
\newcommand{\rhat}{r^{(2)}}
\newcommand{\sbar}{s^{(0)}}

\newcommand{\shat}{s^{(2)}}

\def\barr{\begin{array}}
\def\earr{\end{array}}
\def\half{{\textstyle {1\over 2}}}
\def\third{{\textstyle {1\over 3}}}
\def\fourth{{\textstyle {1\over 4}}}

\jname{}
\jvol{}
\jyear{}
\doi{}

\begin{document}
\hoffset-1in
\voffset -1in
% Title
\title[Stationary Expansion Shocks for a Regularized Boussinesq System]{Stationary Expansion Shocks for a Regularized Boussinesq System}

%Authors, affiliations address.
\author[G.A. El, M.A. Hoefer and M. Shearer]{Gennady El, Mark Hoefer and Michael Shearer\thanks{Address for
correspondence: M. A. Hoefer, Department of Applied Mathematics, University of Colorado,
Boulder, CO 80309, USA; email: hoefer@colorado.edu
 }} \affil{Department of Mathematical Sciences, Loughborough University
Loughborough, UK, LE11 2HN} \affil{Department of Applied Mathematics, University of Colorado,
Boulder, CO 80309, USA} \affil{Department of Mathematics, North Carolina State University,
Raleigh, NC 27695,  USA}

\maketitle

%Abstract
\begin{abstract}
  Stationary expansion shocks have been recently identified as a new
  type of solution to hyperbolic conservation laws regularized by
  non-local dispersive terms that naturally arise in shallow-water
  theory.  These expansion shocks were studied in
  \cite{el_hoefer_shearer_2016} for the Benjamin-Bona-Mahony equation
  using matched asymptotic expansions. In this paper, we extend the
  analysis of \cite{el_hoefer_shearer_2016} to the regularized
  Boussinesq system by using Riemann invariants of the underlying
  dispersionless shallow water equations. The extension for a system
  is non-trivial, requiring a combination of small amplitude,
  long-wave expansions with high order matched asymptotics.  The
  constructed asymptotic solution is shown to be in excellent
  agreement with accurate numerical simulations of the Boussinesq
  system for a range of appropriately smoothed Riemann data.
\end{abstract}

\section{Introduction}
\label{sec:introduction}
We consider  the  normalized form of the classical regularized Boussinesq system for shallow water waves with dispersion 
(see e.g. \cite{whitham_book}, \cite{bona_etal_2002})
   \begin{equation}
  \label{boussinesq1} 
  \barr{rcl}
  h_t+(uh)_x&=&0\\[6pt]
  u_t+uu_x + h_x -\third u_{xxt}&=&0,\earr
\end{equation}
The non-dimensional
variables $h,u$ represent the height of the water free surface above a
flat horizontal bottom, and the depth-averaged horizontal component of
the water velocity, respectively.  System (\ref{boussinesq1}) is non-evolutionary, i.e. not explicitly resolvable with respect to the time derivatives, a property that admits the possibility
%which makes possible the existence
of new classes of solutions 
%principally  not available for 
not generally observed in hyperbolic conservation laws and their
evolutionary dispersive regularizations such as the Korteweg - de
Vries equation, the defocusing nonlinear Schr\"odinger equation,  and
other equations exhibiting rich families of dispersive shock waves
\cite{el_hoefer_dsw_2016,el_hoefer_shearer_mkdv_2016}. New solutions
in the form of stationary, smooth, non-oscillatory expansion shocks
were found in \cite{el_hoefer_shearer_2016} for the
Benjamin-Bona-Mahony (BBM) equation, that represents a uni-directional
analog of the system (\ref{boussinesq1}).

A stationary shock solution of (\ref{boussinesq1}), 
\begin{equation}
  \label{shockb}
  h(x,t)=\left\{\begin{array}{ll}
      h_-, \quad & x<0\\[6pt]
      h_+, &x>0,
    \end{array}  \right. \ \ \ 
  u(x,t)=\left\{\begin{array}{ll}
      u_-, \quad & x<0\\[6pt]
      u_+, &x>0,
    \end{array}
  \right.
\end{equation}
must satisfy Rankine-Hugoniot (RH) jump conditions  
\begin{equation}
  \label{rh2} 
  h_+u_+=h_-u_-; \ \
  h_++\half u_+^2=h_-+\half u_-^2.
\end{equation}
%where $h_\pm = h(\pm \infty,t)$ and $u_\pm = u(\pm \infty,t)$.
Equations \eqref{rh2} can be reformulated to express $u_\pm$ in terms of $h_\pm$ yielding expressions that we refer to as  the {\it RH locus} 
\begin{equation}
  \label{RH_locus}
  u_\pm = h_\mp \left ( \frac{2}{h_- + h_+} \right )^{1/2}.
\end{equation}
Note that (\ref{shockb}), (\ref{rh2}) is a weak solution of both the hyperbolic system of dispersionless shallow-water equations and the  dispersive system (\ref{boussinesq1}), due to the shock being time-independent. The shock is expansive (in the sense specified below) if and only if $h_+<h_-$.  Expansion shocks do not satisfy Lax entropy conditions \cite{lax} and are known to be unstable, immediately giving way to continuous self-similar rarefaction waves in hyperbolic theory.  However, for certain types of dispersive regularization, a smoothed stationary expansion shock  can persist, exhibiting only slow, algebraic decay with time. This new type of shock wave  was identified in the BBM equation \cite{el_hoefer_shearer_2016} by constructing an asymptotic solution of an initial value problem  with smoothed jump (Riemann) initial data.
%stationary expansion shock wave solution of the BBM equation.

In \cite{el_hoefer_shearer_2016}, we also showed numerical simulations of the Boussinesq system (\ref{boussinesq1}) with initial conditions for $h,u$ representing smoothed Riemann data with $h_+<h_-$ satisfying the Rankine-Hugoniot conditions (\ref{rh2}) in the far field. The graphs of the variables $h(x,t),u(x,t)$ as time $t$ evolves resemble the structures observed for the evolution of the asymptotic solution of the BBM equation
%a smoothed step in the BBM equation, for which there is an accurate prediction using matched asymptotic expansions 
\cite{el_hoefer_shearer_2016}. However, the arguments of that paper do not apply directly to the evolution of stationary shocks for the system (\ref{boussinesq1}), and the purpose of this paper is to show how the BBM analysis can be extended to describe the Boussinesq expansion shocks.  It turns out that the generalization of the analysis of \cite{el_hoefer_shearer_2016} to a system requires some subtle manipulations, including expansions with two parameters and a higher order matched asymptotic analysis. The analysis reveals features of the solution  not present in the scalar case.
The central idea is to use Riemann invariants of the underlying dispersionless shallow water system as new field variables in the full dispersive equations (\ref{boussinesq1}). Broadly speaking, the Riemann invariant associated with the faster characteristic speed is constant to a high order, while the Riemann invariant of the slower characteristic speed evolves according to the BBM equation. However, a consistent characterization of this broad behavior requires a careful use of matched asymptotic expansions, with precise control of spatial and temporal scaling, in comparison with the initial jump in the data.  In the final section, we present results of numerical simulations that are in excellent agreement with the asymptotics, over a surprisingly wide range of parameters. Numerical errors are shown to be consistent with the asymptotic predictions,   small inaccuracies being largely explained through higher order terms and wave properties.

%Our analysis reveals contrasting behaviors for the slow and fast Riemann invariants (associated with the slower and faster characteristic speeds, respectively), leading to an asymptotically consistent characterization of both $h(x,t)$ and $u(x,t)$, which is shown to be in  excellent agreement with accurate numerical simulations of the Boussinesq system (\ref{boussinesq1}). 

\section{Expansion shock Riemann data}
The shallow water equations 
 \begin{equation}
  \label{SWE1} 
  \barr{rcl}
  h_t+(uh)_x&=&0\\[6pt]
  u_t+uu_x + h_x  &=&0, \earr
\end{equation}
coincide with the dispersionless limit of the Boussinesq equations (\ref{boussinesq1}).   System (\ref{SWE1}) is a hyperbolic system of conservation laws, with flux function $F(h,u)=(uh,\half u^2+h).$ We shall assume that $u>0.$ The characteristic speeds 
$u\pm \sqrt{h}$
are real and distinct eigenvalues of the Jacobian matrix $dF(h,u),$ for $h>0.$ 
%and represent the speeds $\pm \sqrt{h}$ of small disturbances relative to the underlying flow speed $u.$  
Since we are assuming $u>0,$ we have $\lambda_2=u+\sqrt{h}>0,$ whereas $\lambda_1=u-\sqrt{h}$ can have either sign.
The corresponding Riemann invariants 
\begin{equation}
  \label{eq:2}
     \quad s = u + 2\sqrt{h} , \quad r = u - 2\sqrt{h}
     \end{equation}
diagonalize the system (\ref{SWE1}), which for smooth solutions becomes
\beq\label{SWE2}
s_t+  \fourth(r + 3s)  s_x=0, \qquad   r_t+ \fourth(3 r + s) r_x=0.
\eeq
%Thus on each characteristic $dx/dt=\lambda_\pm,$ the Riemann invariant $r_\pm$ (respectively) is constant, but of course the constant is generally different on different characteristics of the same family. 
Inverse formulae for $h$ and $u$ in terms of the Riemann invariants are
\begin{equation}\label{hu}
h = \tfrac{1}{16} (s-r)^2, \qquad u = \half(s + r). 
\end{equation}
Rarefaction waves for system (\ref{SWE1}) are solutions $h(x,t), u(x,t),$ throughout which one of the Riemann invariants is constant. We will consider only rarefaction waves associated with the slow $\lambda_1=\fourth(3r+s)$ characteristic family; $s$ is constant throughout such a wave, but $r$ is constant only on each individual $\lambda_1$ characteristic. 

As we have mentioned, the stationary shock (\ref{shockb}) is a weak solution of both the Boussinesq system (\ref{boussinesq1}) and of the shallow water equations (\ref{SWE1}). The Lax entropy condition specifies that at each $t>0,$ three of the four characteristics (two for $x>0,$ and two for $x<0$) should enter the shock, and the fourth should leave. Since $\lambda_2>0,$ this is equivalent to requiring $\lambda_1(h_-,u_-)>0,$ and $\lambda_1(h_+,u_+)<0.$ If these inequalities are satisfied, we say the shock is {\em compressive}. If they are reversed, the shock is {\em expansive.} 

We now observe that the stationary shock     (\ref{shockb}) is compressive if and only if $h_+>h_-.$ Correspondingly, it is expansive if and only if $h_+<h_-$. To see this, 
we use  (\ref{RH_locus}) to deduce that $\lambda_1(h_-,u_-)= u_--\sqrt{h_-}=h_+\left ( \frac{2}{h_- + h_+} \right )^{1/2}-\sqrt{h_-}<0$ if and only if $h_+<h_-,$  and similarly, $\lambda_1(h_+,u_+)= u_+-\sqrt{h_+}=h_-\left ( \frac{2}{h_- + h_+} \right )^{1/2}-\sqrt{h_+}>0$ if and only if $h_+<h_-.$ 
\\

We note that the scaling
\begin{equation} \label{jk}
  \tilde{h} = \frac{h}{H}, \quad \tilde{u} = \frac{u}{\sqrt{H}}, \quad
  \tilde{t} = \sqrt{H} t, \quad \tilde{x} = x,
\end{equation}
leaves eq.~\eqref{boussinesq1} invariant. Therefore, without loss of generality, we can consider
\begin{equation}
  \label{eq:27}
  h_+ = 1 , \quad h_-=H.
\end{equation}
Utilizing the normalization \eqref{eq:27} and the RH locus \eqref{RH_locus}, the expansion shock Riemann data  for \eqref{SWE1}, i.e., \eqref{shockb} with $t=0$, become
\begin{equation}
  \label{eq:54}
  h(x,0) =
  \begin{cases}
    H & x < 0, \\
    1 & x > 0,
  \end{cases}
  \quad 
  u(x,0) = \left ( \frac{2}{1 + H} \right )^{1/2} \cdot
  \begin{cases}
    1 & x < 0, \\
    H  & x > 0 .
  \end{cases}, \quad H > 1 ,
\end{equation}
or, equivalently, for \eqref{SWE2},
\begin{equation}
  \label{eq:52}
    r(x,0) =
  \begin{cases}
    \left ( \frac{2}{1+H} \right )^{1/2} - 2 \sqrt{H} & x < 0, \\
    H \left ( \frac{2}{1+H} \right )^{1/2} - 2 & x > 0,
  \end{cases} \quad
  s(x,0) =
  \begin{cases}
    \left ( \frac{2}{1+H} \right )^{1/2} + 2 \sqrt{H} & x < 0, \\
    H \left ( \frac{2}{1+H} \right )^{1/2} + 2 & x > 0 .
  \end{cases}
\end{equation}
In what follows, the initial water height jump parameter,
\begin{equation}
  \label{eq:1}
  \epsilon=2(\sqrt{H}-1),
\end{equation}
plays an important role.  It will be shown in what follows that it is
convenient to utilize the small parameter $\epsilon$ rather than $H-1$
(note that $\epsilon \sim H-1$ for $0 < H-1 \ll 1$) so that the far-field
conditions for $h$ in eq.~\eqref{eq:27} are satisfied exactly in the
obtained approximate solution.  One can see that, if $\epsilon \ll 1$,
then $s(x,0)$ in eq.~\eqref{eq:52} is constant in $x$ to second order in
$\epsilon$,
\begin{equation}
  \label{eq:55}
  s_\pm = 3 + \frac{3}{4}\epsilon + \frac{1}{32}\epsilon^2 +
  \frac{1}{128}\epsilon^3 \cdot
  \begin{cases}
    -3 \\
    1
  \end{cases} + \mathcal{O}(\epsilon^4),
\end{equation}
where $s_{\pm}=s(h_{\pm}, u_{\pm})$, see 
\eqref{shockb}, \eqref{eq:2}. Thus, the initial jump of $s$ across the weak
expansion shock solution is of the third order, $s_+ - s_- =
-\tfrac{1}{32}\epsilon^3 + \mathcal{O}(\epsilon^4)$. At the same time,
the initial jump in $r$,
\begin{equation}
  \label{dr}
  r_{\pm} = -1 + \frac{1}{4} \epsilon \cdot
  \begin{cases}
    3 \\ -5
  \end{cases}
  + \frac{1}{32}\epsilon^2
  + \mathcal{O}(\epsilon^3)
\end{equation}
is of the first order. Thus, for small initial jumps, the RH locus of
the expansion shock coincides to $\mathcal{O}(\epsilon^2)$ with the
simple (rarefaction) wave locus $s=const$. This observation is similar to the well-known property of systems of hyperbolic conservation laws, in which rarefaction curves (for a given constant state) have third order contact with shock curves \cite{lax}. The difference is that in our calculation, both constant states are varied with $H,$ keeping the wave speed constant, whereas in the classical case, the wave speed varies along the wave curves, and one of the constant states is fixed. 

The purpose of using the small parameter $\epsilon$ is due to the fact
that $h_\pm = \frac{1}{16}(r_\pm - s_\pm)^2$ exactly satisfies
\eqref{eq:27}, even for the first and second order expansions in terms
of $\epsilon$ in equations \eqref{eq:55} and \eqref{dr}.  If one
instead expands $s_\pm$, $r_\pm$ in terms of the small parameter
$H-1$, this property will not hold.  Although using $\epsilon$ or
$H-1$ yields asymptotically equivalent approximate solutions, the
sustenance of the far-field behavior in equation \eqref{eq:27} is
useful for comparing the asymptotic solution with the numerical
solution, as we will do in section \ref{sec:numerical-simulation}.

\section{BBM approximation and the structure of the expansion
  shock}\label{sec4}
For hyperbolic conservation laws, Riemann initial data such as in
eq.~\eqref{eq:54} provide useful mathematical approximations to
physical problems in which the data are actually smooth, as well as being the basis for the method of wave front tracking \cite{bressan}.  For the
dispersive problem studied here, the initial transition width turns out to be
an important small parameter in the analysis.  We therefore introduce
\begin{equation}
  \label{eq:30}
  0 < \delta \ll 1,
\end{equation}
as a small 
%(dimensionless) length scale -- this is irrelevant - delta has the same length scale as x, either none or dimension of length 
 scaling parameter  characterizing the width (in $x$) of the transition in the smooth initial data approximating the jump data in equation \eqref{eq:54}. The constants in \eqref{eq:54} now play the 
% smoothed,
%initial transitions between the constant states in eq.~\eqref{eq:54},
%playing now the 
role of far-field data $h_\pm = h(\pm \infty,t)$
and $u_\pm = u(\pm \infty,t)$.  The precise structure of the smooth
transition will be determined in the course of our analysis. To get
some insight into the structure of the evolution of expansion shocks for the Boussinesq
system \eqref{boussinesq1}, we use the proximity of the system
\eqref{boussinesq1} to the BBM equation for the class of Riemann data
\eqref{eq:54} with small jumps, $0 < \epsilon = 2(\sqrt{H}-1)
\ll 1$.  To this end, we convert 
%introduce the Riemann variables \eqref{eq:2} in 
the full dispersive system \eqref{boussinesq1} to Riemann invariant variables \eqref{eq:2}, resulting in the system 
%exactly obtain
\begin{equation}
  \label{eq:3}
  \begin{split}
    r_t + \fourth(3 r + s) r_x &= \tfrac{1}{6} ( r_{xxt} + s_{xxt})
    \\
    s_t + \fourth(r + 3s) s_x &= \tfrac{1}{6} ( r_{xxt} + s_{xxt}) .
  \end{split}
\end{equation}
A similar change of variables to \eqref{eq:2}, \eqref{hu} was
previously used in a fully nonlinear model of shallow
capillary-gravity waves, the generalized Serre system, in order to
obtain approximate unidirectional models, splitting the slow and fast
waves \cite{milewski}.  Here, we demonstrate the utility of these
variables for obtaining approximate solutions to the original
bi-directional model.

Motivated by the Riemann data expansions \eqref{eq:52}, \eqref{eq:55}
for small jumps, we consider initial data for the Boussinesq system
\eqref{eq:3} with constant $s=\sbar$. Then, having initially $s=\sbar$ and $r$ exhibiting a jump, we can neglect $s_{xxt}$ in the first
equation of \eqref{eq:3}, at least for $t \ll1$, and reduce it to the
BBM equation
\begin{equation}\label{BBM}
  v_t+vv_x=\tfrac16 v_{xxt} \, ,
\end{equation}
provided
\begin{equation}\label{v}
v=\tfrac14 (\sbar + 3r).
\end{equation}
Then, if $v(x,0)= A \tanh (x/\delta)$, the approximate solution for
the expansion shock of the BBM equation \eqref{BBM} is available from
\cite{el_hoefer_shearer_2016}. The related behaviors of $h(x,t),
u(x,t)$ are then found from \eqref{hu}.  The described BBM
approximation, while not yet being fully justified asymptotically,
provides some useful intuition into the expansion shock structure for
the Boussinesq system and in fact, as we shall see, correctly
describes the first order asymptotic solution.

\begin{figure}
  \centering
  \includegraphics[width=0.8\columnwidth]{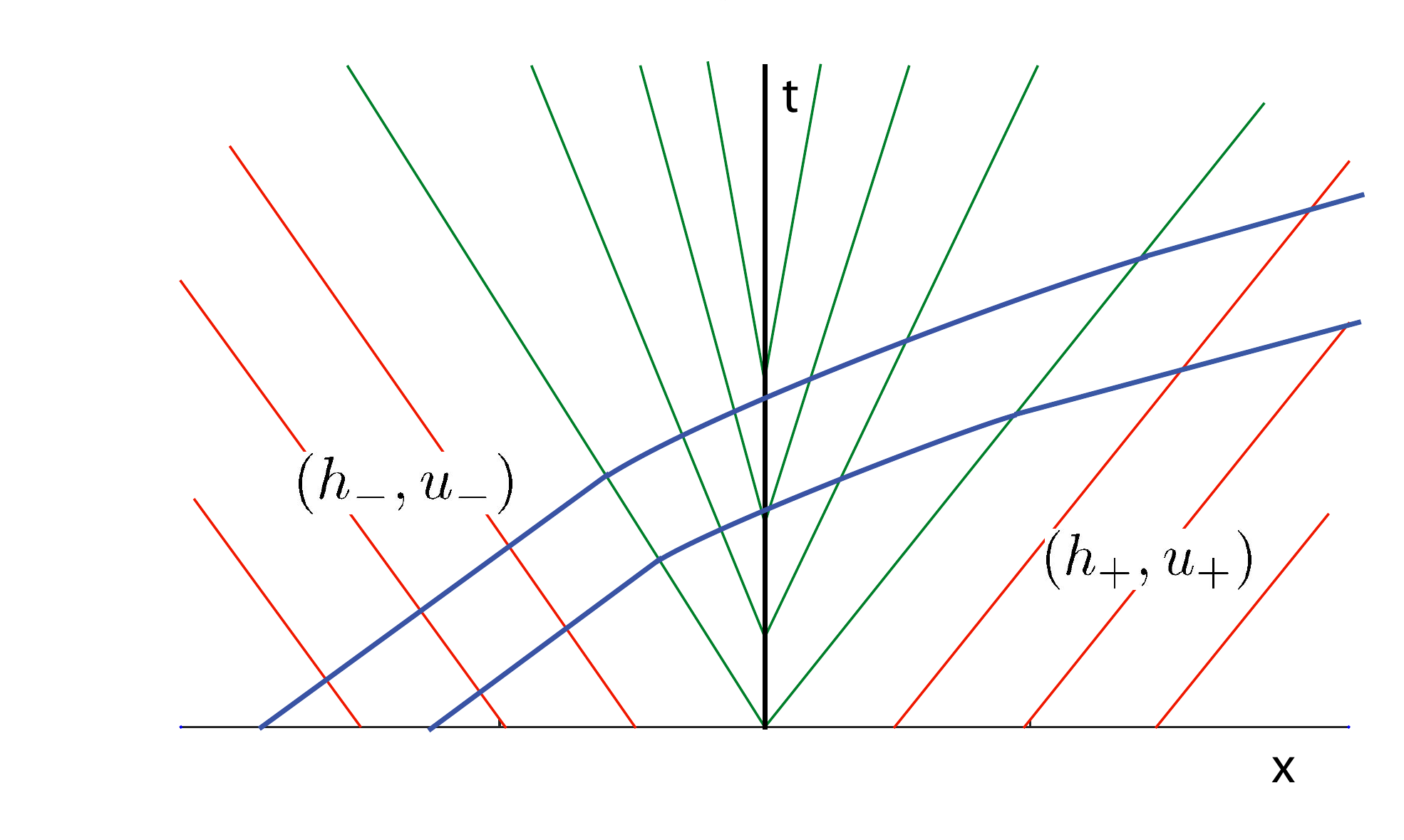}
  \caption{Characteristics in the $(x,t)$-plane for the expansion
    shock solution of system (\ref{boussinesq1}).      \label{exp_characteristics}}
\end{figure}
In Fig.~\ref{exp_characteristics}, we show schematically the structure
of characteristics for the evolution of the Boussinesq stationary
expansion shock suggested by the BBM approximation.  The smoothed jump
initial data for $h,u$ are indicated as subscripts ``$-$'' for $-x \gg
\delta,$ and ``$+$'' for $+x \gg \delta$. In the rarefaction wave,
the characteristic speed $\lambda_1$ is varying, increasing from left
to right, and the characteristics have correspondingly different
speeds as they emerge from the shock. Moreover, $\lambda_2$ is
changing also, so the fast characteristics are curved as they pass
through the simple wave.    But the Riemann invariant $s$ is
constant on these characteristics, and therefore, takes the value
$s_-=s(h_-,u_-)$ to the left, and $s_+=s(h_+,u_+)$ to the right.  As
we have shown (see \eqref{eq:55}), the jump $s_+-s_-$ across $x=0$ is
small for small to moderate water height jumps $\epsilon$.  As a
consequence, $s$ turns out to be roughly constant across the entire
solution except, as we shall see, in the small, $\delta$-wide region
within the expansion shock.

\section{Expansion shock for the Boussinesq equations}

We now proceed with the detailed asymptotic analysis of expansion
shocks for the system \eqref{eq:3}. Similar to
\cite{el_hoefer_shearer_2016}, we use matched asymptotic expansions.
The key to the analytic construction in \cite{el_hoefer_shearer_2016}
is the separable structure of the PDE describing the inner solution
with the scaled variables $\xi=x/\delta$, $T=\delta t$.
Unfortunately, the %unapproximated
Boussinesq equations \eqref{eq:3} do not admit such a separation of
variables with this scaling of $x,t$ and require a somewhat more
sophisticated asymptotic analysis to reveal the detailed internal
structure of the expansion shock.

We first consider the inner problem, i.e., near the initial smoothed
transition. The precise structure of the smoothed Riemann data will be
clarified in the analysis below. Our construction will be based on
formal expansions in two small parameters: the initial jump in height
$\epsilon \sim H-1$ (eq.~\eqref{eq:1}) and the jump spatial transition
width $\delta$, set by the initial conditions.  However, as we shall
see, the resulting solution also provides an excellent approximation
for moderate values of $H-1$.

\subsection{Inner Solution: first order approximation}
\label{sec:inner-solution}
Assuming the spatial scale $\delta$ of the inner solution to be
set by the smoothed initial data, we seek a solution to
eqs.~\eqref{eq:3} in the scaled inner variables $\xi = x/\delta$,
$\tau = \mu t$, where the parameter $0 < \mu \ll 1$ is an inverse
timescale for the development of an expansion shock, which is to be
determined. With these scalings, equations \eqref{eq:3} become
\begin{equation}
  \label{eq:39}
  \begin{split}
    \mu \ri_\tau + \frac{1}{4 \delta}(3\ri + \si)\ri_\xi &=
    \frac{\mu}{6\delta^2} ( \ri_{\xi\xi\tau} +
    \si_{\xi\xi\tau}) \\
    \mu \si_\tau + \frac{1}{4 \delta}(\ri + 3\si)\si_\xi &=
    \frac{\mu}{6\delta^2} ( \ri_{\xi\xi\tau} + \si_{\xi\xi\tau}) .
  \end{split}
\end{equation}
We wish to find an approximate solution of system~\eqref{eq:39} that
agrees with the initial conditions \eqref{eq:52} to some order of
accuracy, specifically for $t=\tau=0,$ and $\xi\to \pm\infty.$  For this, we assume that the parameter $\epsilon$ in \eqref{eq:1} is small
and expand $\ri$ and $\si$ according to
\begin{equation}
  \label{eq:40}
  \begin{split}
    \ri(\xi,\tau) &= \rbar + \epsilon \rtil(\xi,\tau) +
    \epsilon^2 \rhat(\xi,\tau) + \cdots, \\
    s &= \sbar + \epsilon^2 \shat(\xi,\tau) +  \cdots,
  \end{split}
\end{equation}
where $\rbar$, $\rtil$, $\rhat$, $\sbar$, and $\shat$ are
$\mathcal{O}(1)$ as $\epsilon \to 0$, $\delta \to 0$, $\mu \to 0$.
The parameter $\epsilon$ is proportional to the initial jump in $r$
from eq.~\eqref{dr}, and in the expansion for $s$ we have assumed that
$s^{(1)}=0,$ which is consistent with the discussion in the previous
section, and could be readily deduced by modifying the analysis below.

Inserting expansions \eqref{eq:40} into eq.~\eqref{eq:39}, we obtain
\begin{equation}
  \label{eq:29}
  \begin{split}
    &\mu \left ( \epsilon \rtil_\tau + \cdots \right ) \\
    & +\frac{1}{4
      \delta} \left ( 3 \rbar + \sbar + 3\epsilon \rtil +
      \epsilon^2 ( 3 \rhat + \shat ) + \cdots \right ) ( \epsilon
    \rtil_\xi +
    \epsilon^2 \rhat_\xi + \cdots ) \\
    &= \frac{\mu}{6 \delta^2} \left ( \epsilon \rtil_{\xi\xi \tau}
      + \epsilon^2(\rhat_{\xi\xi\tau} + \shat_{\xi\xi\tau}) + \cdots
    \right )
  \end{split}
\end{equation}
and
\begin{equation}
  \label{eq:41}
  \begin{split}
    \left ( \mu \epsilon^2 \shat_\tau + \cdots
      \right ) &+ \frac{1}{4 \delta} \left ( \rbar + 3\sbar +
        \cdots \right ) ( \epsilon^2 \shat_\xi + \cdots
       ) \\
       &= \frac{\mu}{6 \delta^2} ( \epsilon
    \rtil_{\xi\xi \tau} + \cdots ) .
  \end{split}
\end{equation}
If we assume $\mu \ll 1 / \delta$, then the leading order term
in eq.~\eqref{eq:29} is
\begin{equation}
  \label{eq:42}
  \mathcal{O}\left ( \frac{\epsilon}{\delta} \right ): \quad
  \frac{1}{4}(3 \rbar + \sbar )\rtil_\xi = 0 ,
\end{equation}
which is solved by
\begin{equation}
  \label{eq:43}
  \rbar = -\frac{\sbar}{3} .
\end{equation}
In order to find an approximate inner solution that balances
nonlinearity and dispersion in eq.~\eqref{eq:29}, we require  
\begin{equation}
  \label{eq:4}
  \mu \epsilon \ll \epsilon^2/\delta = \mathcal{O}(\mu
\epsilon/\delta^2).
\end{equation}
This determines the inverse timescale $\mu$ in terms of the transition width
$\delta$ and jump amplitude parameter $\epsilon$.  We take
\begin{equation}
  \label{eq:7}
  \mu = \delta \epsilon .
\end{equation}

Proceeding under these assumptions, we obtain from eq.~\eqref{eq:29}
\begin{equation}
  \label{eq:45}
  \mathcal{O}\left ( \frac{\epsilon^2}{\delta} \right ): \quad
  \rtil \rtil_\xi = \frac{2}{9}
  \rtil_{\xi\xi\tau} .
\end{equation}
We can now solve this equation by separation of variables
\begin{equation}
  \label{eq:48}
  \rtil(\xi,\tau) = a(\tau) f(\xi) 
\end{equation}
where
\begin{align}
  \label{eq:9}
  \frac{2 \dot{a}}{9 a^2} &= \frac{f f'}{f''} = -K,
\end{align}
and $K > 0$ is the separation constant.  We determine $a$ and $f$ as
for BBM in \cite{el_hoefer_shearer_2016}
\begin{equation}
  \label{eq:11}
  a(\tau) = \frac{A}{\frac{9}{2} A K \tau + 1}, \quad f(\xi) = B
  \tanh\left ( \frac{B}{2K} \xi \right ) ,
\end{equation}
where $A > 0$ and $B > 0$ are parameters to be determined.  
%%% Not sure why this was included.
% Then
% \begin{equation}
%   \label{eq:12}
%   \begin{split}
%     \frac{1}{4} \dot{a}(\tau) = - \frac{9 K A^2}{8
%       \left (
%         \frac{9}{2} A K \tau + 1 \right )^2}.
%   \end{split}
% \end{equation}
We choose the parameters
\begin{equation}
  \label{eq:13}
  B = 1, \quad K = \frac{1}{2},
\end{equation}
and retain the amplitude parameter $A$, which will be determined by
the RH locus, so that the solution \eqref{eq:11} is
\begin{equation}
  \label{eq:23}
  a(\tau) = \frac{A}{\frac{9}{4}A \tau + 1}, \quad f(\xi) = \tanh(\xi) .
\end{equation}
Then,  the approximate inner expansion shock solution to first order in
$\epsilon$ can be written
\begin{align}
  \label{eq:14}
  \ri(\xi,\tau) &= - \frac{\sbar}{3} + \frac{ \epsilon A }{\frac{9}{4}
    A\tau + 1}\tanh(\xi) + \mathcal{O}(\epsilon^2), \\
  \label{eq:15}
  \si(\xi,\tau) &= \sbar + \mathcal{O}(\epsilon^2). 
\end{align}

In order to determine the free parameters $\sbar$ and $A$ in terms
of the initial data, we evaluate the solution \eqref{eq:14},
\eqref{eq:15} at $t=0$, $\xi \to \pm \infty$ and compare it with the
first order small-jump expansions of the initial conditions
\eqref{eq:52} incorporating the RH locus:
\begin{align}
  \label{eq:18a}  
  r_{\pm} &= -\frac{\sbar}{3} \pm \epsilon A + \mathcal{O}(\epsilon^2), \\
  \label{eq:18b}
  s_{\pm} &= \sbar + \mathcal{O}(\epsilon^2), \quad \xi \to \pm \infty .
\end{align}
Comparing eqs.~\eqref{eq:18a} and \eqref{eq:18b} with
eqs.~\eqref{eq:55} and \eqref{dr}, we find
\begin{equation}
  \label{eq:5}
  \sbar = 3 + \frac{3}{4} \epsilon, \quad A = 1 .
\end{equation}

We note that the constructed first order inner solution \eqref{eq:14},
\eqref{eq:15}, \eqref{eq:5} for the expansion shock simultaneously
incorporates the simple wave locus $s_-=s_+=\sbar$ of the shallow
water equations \eqref{SWE2} {\it and} the RH condition $v_-+v_+ =0$
for the stationary shock of the simple wave equation $v_t+vv_x=0$,
where $v=\tfrac14(\sbar +3r)$ (recall eq.~\eqref{v}). This is nothing
but the dispersionless limit of the BBM equation \eqref{BBM}. Indeed,
one can see that the first order solution written in terms of $v$
agrees with the inner solution for the BBM expansion shock obtained in
\cite{el_hoefer_shearer_2016}.  We also note that, within the first
order approximation, the ordering between the small parameters
$\epsilon$ and $\delta$ is $\mu = \delta \epsilon \ll 1/\delta$ or
$\epsilon \ll 1/\delta^2$.

\subsection{Inner solution: second order approximation}
\label{sec:second-order-solut}

To obtain the $\mathcal{O}(\epsilon^2)$ correction, we consider
equation \eqref{eq:41}, from which we deduce, using $\rtil=a(\tau)f(\xi)$ from the previous subsection,
\begin{equation}
  \label{eq:46}
  \mathcal{O}\left ( \frac{\epsilon^2}{\delta} \right ): \quad
  \shat_\xi = \frac{1}{12}
  \rtil_{\xi\xi\tau} = \frac{1}{12} \dot{a}(\tau) f''(\xi).
\end{equation}
This equation  is solved with
\begin{equation}
  \label{eq:8}
  \shat(\xi,\tau) = \frac{1}{12} \dot{a}(\tau)f'(\xi) + C =
  -\frac{3
    \,\mathrm{sech}^2(\xi)}{16 \left ( 1+ \frac{9}{4} \tau
    \right )^2} + C. 
\end{equation}
The constant of integration $C$ could at this stage be a function of $\tau,$ but it will be determined below by matching to the far field, so it is necessarily constant.

We now proceed to the next order equation in \eqref{eq:29}, assuming
that $\delta \epsilon^2 \ll \epsilon^3/\delta$, implying the basic
small parameter ordering
\begin{equation} 
  \label{eq:64}
  \delta \ll \epsilon^{1/2} .
\end{equation}
We find the equation for $\rhat:$
\begin{equation}
  \label{eq:32}
  \mathcal{O}\left ( \frac{\epsilon^3}{\delta} \right ): \quad
  \rhat_{\xi\xi\tau} - \frac{9}{2} (\rtil \rhat)_\xi =
  \frac{3}{2} 
  \shat \rtil_{\xi} - \shat_{\xi\xi\tau} .
\end{equation}
We observe that this equation has solutions of the form% can be solved by separation of variables according to
\begin{equation}
  \label{eq:34}
  \rhat(\xi,\tau) = a^2(\tau) g(\xi) - \frac{C}{3},
\end{equation}
in which $a(\tau)$ is given in \eqref{eq:23}.
Then $g$ satisfies
\begin{equation}
  \label{eq:36}
  g'' + (fg)' = \frac{1}{16} \left ( f'^2 + 3 f''' \right ) = 
   \frac{1}{16} \left ( f' - f' f^2 + 3 f''' \right ) .
\end{equation}
Integrating, we obtain
\begin{equation}
  \label{eq:37}
  g' + fg = \frac{1}{16} \left ( f - \frac{1}{3} f^3 + 3 f'' + \frac{1}{3}D
  \right ),
\end{equation}
where $D$ is a constant of integration.  An integrating factor for
this equation is $\mathrm{cosh} (\xi)$.  We therefore obtain
\begin{equation}
  \label{eq:38}
  g(\xi) = \frac{1}{16} \left ( \frac{2}{3} + \frac{17}{3}
    \,\mathrm{sech}^2(\xi) + \frac{1}{3}D \tanh(\xi) + \frac{1}{3}E\,\mathrm{sech}(\xi) \right ) ,
  % g(\xi) = \frac{1}{48} \left ( 2 + 17 \, \mathrm{sech}^2 (\xi)
  %   + D\,\tanh (\xi) + E\, \mathrm{sech} (\xi) \right ).
\end{equation}
where $E$ is an additional constant of integration.  Then the
approximate inner solution for the expansion shock to second order in
$\epsilon$ becomes
\begin{equation}
  \label{eq:50}
  \begin{split}
    \ri(\xi,\tau) &\sim -1 + \epsilon \left ( - \frac{1}{4} +
      \frac{\tanh(\xi)}{1 + \frac{9}{4} \tau} \right )  \\
    &\qquad ~~ + \frac{\epsilon^2}{3} \left ( -C + \frac{ 2 + 17 \,
        \mathrm{sech}^2 (\xi) + D \tanh(\xi) + E\,\mathrm{sech}
        (\xi)}{16 \left (
          1 + \frac{9}{4} \tau \right)^2} \right ) , \\
    \si(\xi,\tau) &\sim 3 + \frac{3}{4} \epsilon + \epsilon^2 \left ( C -
      \frac{3\,\mathrm{sech}^2(\xi)}{16 \left (1 +  \frac{9}{4} \tau
        \right )^2} \right ). 
  \end{split}
\end{equation}
  The constant $E$ is a free parameter, not determined at
this order.  We therefore set $E = 0$.  To determine the remaining
parameters $C$ and $D$, we invoke the smoothed Riemann data
\eqref{eq:52} and evaluate the approximate solution \eqref{eq:50} at
$t = 0$ for $\ri$ and $\si$ as $\xi \to \pm \infty$, yielding
(cf.~\eqref{eq:18a}, \eqref{eq:18b}),
\begin{equation}
  \label{eq:33}
  \begin{split}
    r_{\pm} &= -1 + \epsilon \left ( -\frac{1}{4} \pm 1 \right
    ) + \epsilon^2 \left ( -\frac{C}{3} +\frac{1}{24} \pm
      \frac{D}{48}\right ) +
    \mathcal{O}(\epsilon^3) , \\
    s_\pm &= 3 + \frac{3}{4} \epsilon + C \epsilon^2 + \mathcal{O}(\epsilon^3),
  \end{split}
\end{equation}
which satisfy the RH locus expansions \eqref{eq:55} and \eqref{dr} to
$\mathcal{O}(\epsilon^2)$ if we take
\begin{equation}
  \label{sbar}
  C = \frac{1}{32}, \quad D = 0,
\end{equation}  
which, together with \eqref{eq:50}, fully defines the second order
inner solution, beyond the BBM approximation as
\begin{equation}
  \label{eq:18}
  \begin{split}
    \ri(\xi,\tau) &\sim -1 + \epsilon \left ( - \frac{1}{4} +
      \frac{\tanh(\xi)}{1 + \frac{9}{4} \tau} \right )  
    %&\qquad ~~ 
    + \frac{\epsilon^2}{48} \left ( -\frac{1}{2} + \frac{ 2 + 17 \,
        \mathrm{sech}^2 (\xi)}{\left (1 +
          \frac{9}{4} \tau \right)^2} \right ) , \\
    \si(\xi,\tau) &\sim 3 + \frac{3}{4} \epsilon + \frac{\epsilon^2}{16}
    \left ( \frac{1}{2} - \frac{3\,\mathrm{sech}^2(\xi)}{\left (
          1 + \frac{9}{4} \tau \right )^2} \right ) .
  \end{split}
\end{equation}

%%  Very confusing to me and not worth it in my view (MAH). 
% In concluding this section we remark that the second order correction
% \eqref{eq:8} for $s$ actually comes from the same order $\mathcal
% {O}(\epsilon^2/\eps)$ equation in the expansion \eqref{eq:41} as the
% first order correction for $r$ comes in \eqref{eq:29}. The necessity to move
% to the higher order $\mathcal{O}(\epsilon^3/\eps)$ equation in
% \eqref{eq:29} comes from the requirement to have the same
% $\mathcal{O}(\epsilon^2)$ accuracy in $r$ and $s$ so as to consistently
% reconstruct $h$ and $u$ from \eqref{hu}.

\subsection{Outer Solution}
\label{sec:outer-solution}

For matching purposes, it is natural to set the timescale for the
outer scaling to be the same as the timescale of the inner scaling
$\tau = \mu t = \delta \epsilon t$, using \eqref{eq:7}.  Along with
that, we use the long wave, hydrodynamic scaling $X = \delta x$, which
is independent of the jump amplitude parameter $\epsilon$. Then the
leading order (in $\delta$) equations from \eqref{eq:3} are the
dispersionless shallow water equations
\begin{equation}
  \label{eq:21}
  \begin{split}
    \epsilon \ro_\tau + \frac{1}{4}(3 \ro + \so) \ro_X &=  0,
    \\
   \epsilon  \so_\tau + \frac{1}{4}(\ro + 3\so) \so_X &= 0 .
  \end{split}
\end{equation}
We expect a simple wave solution, which we expand as
\begin{equation}
  \label{eq:17}
  \begin{split}
    \so(X,\tau) &= 3 + \frac{3}{4} \epsilon + \frac{1}{32} \epsilon^2 +
    \cdots, \\
    \ro(X,\tau) &= -1 + \epsilon \left ( -\frac{1}{4} + r_1(X,\tau) \right ) +
    \epsilon^2 \left (-\frac{1}{96} + r_2(X,\tau) \right ) + \cdots .
  \end{split}
\end{equation}
With these expansions, the equation \eqref{eq:21} for $\so$ is
identically satisfied.  The equation for $\ro$, expanded in powers of
$\epsilon$, yields to leading order
\begin{equation}
  \label{eq:49}
  \mathcal{O}(\epsilon^2): \quad r_{1,\tau} + \frac{3}{4} r_1 r_{1,X} = 0 , 
\end{equation}
which can be solved with
%  and each order solved.  We take
% an alternative approach and determine a simple wave solution in the
% form
\begin{equation}
  \label{eq:26}
  r_1(X,\tau) = F_1^{(\mathrm{sgn}\,X)} \left (\tau - \frac{4X}{3r_1}
  \right ) ,  
\end{equation}
where we use the functions $F_1^{(\pm)}$ depending on whether $\pm X >
0$.  Matching this to the inner solution \eqref{eq:18} at
$\mathcal{O}(\epsilon)$ yields
\begin{equation}
  \label{eq:53}
  \lim_{X \to 0^\pm} r_1(X,\tau) = F_1^{(\pm)}(\tau) = \lim_{\xi \to
    \pm \infty} \rtil(\xi,\tau) = \pm \frac{1}{1 +
      \frac{9}{4} \tau} 
\end{equation}
Then
\begin{equation}
  \label{eq:31}
  r_1 = \frac{\mathrm{sgn}\,X}{1 + \frac{9}{4} \tau -
      \frac{3X}{r_1}},
\end{equation}
which is solved by
\begin{equation}
  \label{eq:35}
  r_1(X,\tau) = \frac{\mathrm{sgn}\, X + 3X}{1 + \frac{9}{4}
      \tau }  .
\end{equation}

Proceeding to the next order in the expansion of eq.~\eqref{eq:21}
yields an equation for $r_2$
\begin{equation}
  \label{eq:65}
  \mathcal{O}(\epsilon^3): \quad r_{2,\tau} + \frac{3}{4} (r_1 r_2)_X =
  0 .
\end{equation}
One can verify by direct substitution that 
\begin{equation}
  \label{eq:66}
  r_2(X,\tau) = F_2^{(\mathrm{sgn}\,X)}\frac{1+3|X|}{\left (1 +
      \frac{9}{4} \tau \right)^2},
\end{equation}
solves eq.~\eqref{eq:65}.  Matching to the inner solution
\eqref{eq:18}, we obtain
\begin{equation}
  \label{eq:67}
  \lim_{X \to 0^\pm} r_2(X,\tau) = F_2^\pm \frac{1}{\left (1 +
      \frac{9}{4} \tau \right )^2} = \lim_{\xi \to \pm \infty}
  \rhat(\xi,\tau) = \frac{1}{24 \left ( 1+\frac{9}{4}
      \tau \right )^2} ,
\end{equation}
so that $F_2^+ = F_2^- = 1/24$, yielding the second order correction
to the outer solution
\begin{equation}
  \label{eq:68}
  r_2(X,\tau) = \frac{1 + 3|X|}{24\left (1+ \frac{9}{4} \tau
    \right )^2} .
\end{equation}
The approximate outer solution therefore has the form
\begin{equation}
  \label{eq:58}
  \begin{split}
    \ro(X,\tau) &= -1 + \epsilon \left ( -\frac{1}{4} +
      \frac{\mathrm{sgn}\, X + 3X}{1+\frac{9}{4} \tau} \right ) 
    \\
    &\quad \, + \frac{\epsilon^2}{24} \left ( -\frac{1}{4} + \frac{
        1 + 3|X|}{\left (1 + \frac{9}{4} \tau \right )^2} \right ) +
    \mathcal{O}(\epsilon^3), \\
    \so(X,\tau) &= 3 + \frac{3}{4} \epsilon + \frac{1}{32} \epsilon^2 +
    \mathcal{O}(\epsilon^3) .
  \end{split}
\end{equation}
Note that for the dispersionless eq.~\eqref{eq:21} to be a valid
asymptotic approximation of the full Boussinesq eqs.~\eqref{eq:3} to
$\mathcal{O}(\epsilon^3)$, we require the dispersive term to be
negligible to the order considered, i.e., $\epsilon \delta^3 \ll
\epsilon^3$ or $\delta \ll \epsilon^{2/3}$.  This is a less stringent
condition on scale separation than the restriction \eqref{eq:64}
applied for the calculation of the inner solution.

This approximate outer solution is only valid within an expanding
region.  We invoke continuous matching to the far-field along the
two lines
\begin{equation}
  \label{eq:57}
  X = c_\pm \tau ,
\end{equation}
where $c_\pm$ is determined by the requirement
\begin{equation}
  \label{eq:59}
  \ro(c_\pm \tau,\tau) = r_\pm .
\end{equation}
A calculation using \eqref{dr}, \eqref{eq:58} yields the speeds
\begin{equation}
  \label{eq:60}
  c_\pm = \pm \frac{3}{4} + \frac{1}{32}\epsilon
  +  \cdots .
\end{equation}
Matching to the far-field, we obtain the approximate, piecewise smooth
outer solution
\begin{equation}
  \label{eq:25}
  \begin{split}
    \scriptstyle
    \ro(X,\tau) &= 
    \scriptstyle -1 + \epsilon 
      \begin{cases}
        \textstyle - \frac{5}{4} + \frac{1}{32} \epsilon &
        \scriptstyle
        \frac{X}{\tau} \le c_-, \\[3mm]
        \textstyle - \frac{1}{4} + \frac{\mathrm{sgn}\, X + 3X}{1 +
          \frac{9}{4} \tau}
        + \frac{\epsilon}{24} \left ( -\frac{1}{4} + \frac{1+3|X|}{\left
            ( 1 + \frac{9}{4}\tau \right )^2} \right ) 
        & 
        \scriptstyle
        c_- <  \frac{X}{\tau} < c_+, \\[3mm]
        \textstyle \frac{3}{4} + \frac{1}{32}\epsilon & 
        \scriptstyle
        c_+ \le \frac{X}{\tau},
      \end{cases}  +  \mathcal{O}(\epsilon^3) , \\
    \so(X,\tau) &= 3 + \frac{3}{4} \epsilon + \frac{1}{32} \epsilon^2 +
    \mathcal{O}(\epsilon^3). 
  \end{split}
\end{equation}

\subsection{Uniformly Valid Asymptotic Solution}
\label{sec:unif-valid-asympt}

In order to construct a uniformly valid (in $x$) asymptotic solution
to $\mathcal{O}(\epsilon^2)$ in $r$ and $s$, we introduce the composite
solution
\begin{equation}  \label{eq:10}
  %\begin{split}
  \barr{rcl}
    r(x,t) &=& \ri(x/\delta,\delta \epsilon t) + \ro(\delta
    x,\delta \epsilon t) -
    r^{(\mathrm{overlap})}(x,t;\delta,\epsilon), \\ 
    s(x,t) &= &\si(x/\delta,\delta \epsilon t) + \so(\delta
    x,\delta \epsilon t) -
    s^{(\mathrm{overlap})}(x,t;\delta,\epsilon) .
    \earr
  %\end{split}
\end{equation}
We subtract the ``overlap'' portion (common to both the inner and
outer solutions) so that we do not double count the matching region.
We therefore have
\begin{equation}
  \label{eq:16}
  \begin{split}
    r^{(\mathrm{overlap})}(x,t;\delta,\epsilon) &\sim -1 +
    \epsilon \left ( -\frac{1}{4} +
      \frac{\mathrm{sgn}\,X}{1 + \frac{9}{4} \tau} \right ) +
    \frac{\epsilon^2}{24} \left ( -\frac{1}{4} +
      \frac{1}{ \left (1 + \frac{9}{4} \tau \right
        )^2} \right ), \\
    s^{(\mathrm{overlap})}(x,t;\delta,\epsilon) &\sim 3 + \frac{3}{4}
    \epsilon + \frac{1}{32} \epsilon^2 .
  \end{split}
\end{equation}
Then the uniformly valid, composite asymptotic solution for an
expansion shock is
\begin{equation}
  \label{eq:22}
  \begin{split}
    r(x,t;\delta,\epsilon) &\sim -1 + \epsilon \left ( \frac{\tanh\left
          ( \frac{x}{\delta}\right) - \,\mathrm{sgn}\, (x)}{1 +
        \frac{9}{4} \delta \epsilon t} +
      G_1(x,t;\delta,\epsilon) \right ) \\
    &\qquad + \epsilon^2 \left ( \frac{17 \,\mathrm{sech}^2\left (
          \frac{x}{\delta} \right )}{48 \left ( 1+ \frac{9}{4}
          \delta \epsilon t \right )^2} +
      G_2(x,t;\delta,\epsilon) \right ), \\[3mm]
    s(x,t) &\sim 3 + \frac{3}{4} \epsilon + \frac{\epsilon^2}{16} \left (
      \frac{1}{2} - \frac{3 \, \mathrm{sech}^2\left
          (\frac{x}{\delta} \right )}{ \left ( 1 + \frac{9}{4}
          \delta \epsilon t \right )^2}
    \right ), \\[3mm]
    G_1(x,t;\delta,\epsilon) &= \begin{cases} \displaystyle -
      \frac{5}{4} & \displaystyle
      \frac{x}{\epsilon t} \le c_-, \\[3mm]
      \displaystyle -\frac{1}{4} + \frac{\mathrm{sgn}\,(x) + 3\delta
        x}{1+ \frac{9}{4} \delta \epsilon t} &
      \displaystyle
      c_- <  \frac{x}{\epsilon t} < c_+, \\[3mm]
      \displaystyle \frac{3}{4} & \displaystyle  c_+ \le
      \frac{x}{\epsilon t},
    \end{cases} \\[3mm]
    G_2(x,t;\delta,\epsilon) &= \begin{cases} \displaystyle
      \frac{1}{32} & \displaystyle 
      \frac{x}{\epsilon t} \le c_-, \\[3mm]
      \displaystyle -\frac{1}{96}
      + \frac{1+3\delta |x|}{24 \left ( 1 +
            \frac{9}{4} \delta \epsilon t \right )^2} & 
      \displaystyle
      c_- <  \frac{x}{\epsilon t} < c_+, \\[3mm]
      \displaystyle  \frac{1}{32} & 
      \displaystyle
       c_+ \le \frac{x}{\epsilon t},
      \end{cases}
  \end{split}
\end{equation}
where $c_\pm$ is given in \eqref{eq:60}.  This solution can be used in
\eqref{eq:2} to reconstruct the expansion shock water height $h$ and
horizontal velocity $u$.

\section{Numerical Simulation}
\label{sec:numerical-simulation}

We validate the asymptotic analysis of \S\ref{sec:unif-valid-asympt}
with direct numerical simulations of the Boussinesq equations
\eqref{boussinesq1} with initial data consisting of the approximate
expansion shock solution \eqref{eq:22} evaluated at $t = 0$.  The
numerical method is described in the appendix.

Figure \ref{fig:numerics_h_u} depicts the numerical evolution of $h$
and $u$ with an initial jump in $h$ from unity to $H = 1.4$ and the
transition width $\delta = 0.1$.  The boundary conditions $u_\pm$,
determined by eqs.~\eqref{hu}, \eqref{eq:55}, and \eqref{dr}, satisfy
the RH locus \eqref{RH_locus} to order $\epsilon^2$.  The sharp
initial step evolves into an expansion shock that algebraically decays
between non-centered rarefaction waves propagating left and right.
The uniform asymptotic approximation \eqref{eq:22} closely follows the
numerical solution; the most noticeable deviations occurring at the
weak discontinuities, where the rarefactions meet the far-field
boundary conditions with a jump in the first derivative of the
asymptotic solution.  A close examination reveals the generation of a
small amplitude dispersive wavepacket that propagates to the right
(see insets at $t = 15$).  This is due to the fact that the initial
data only approximately corresponds to an expansion shock, accurate to
order $\epsilon^2$.  For this simulation, $\epsilon = 2(\sqrt{H}-1)
\approx 0.366,$ for which $\epsilon^3\sim 0.05,$ larger than the size
of the dispersive wavepacket.
\begin{figure}
  \centering
  \includegraphics[scale=0.33333]{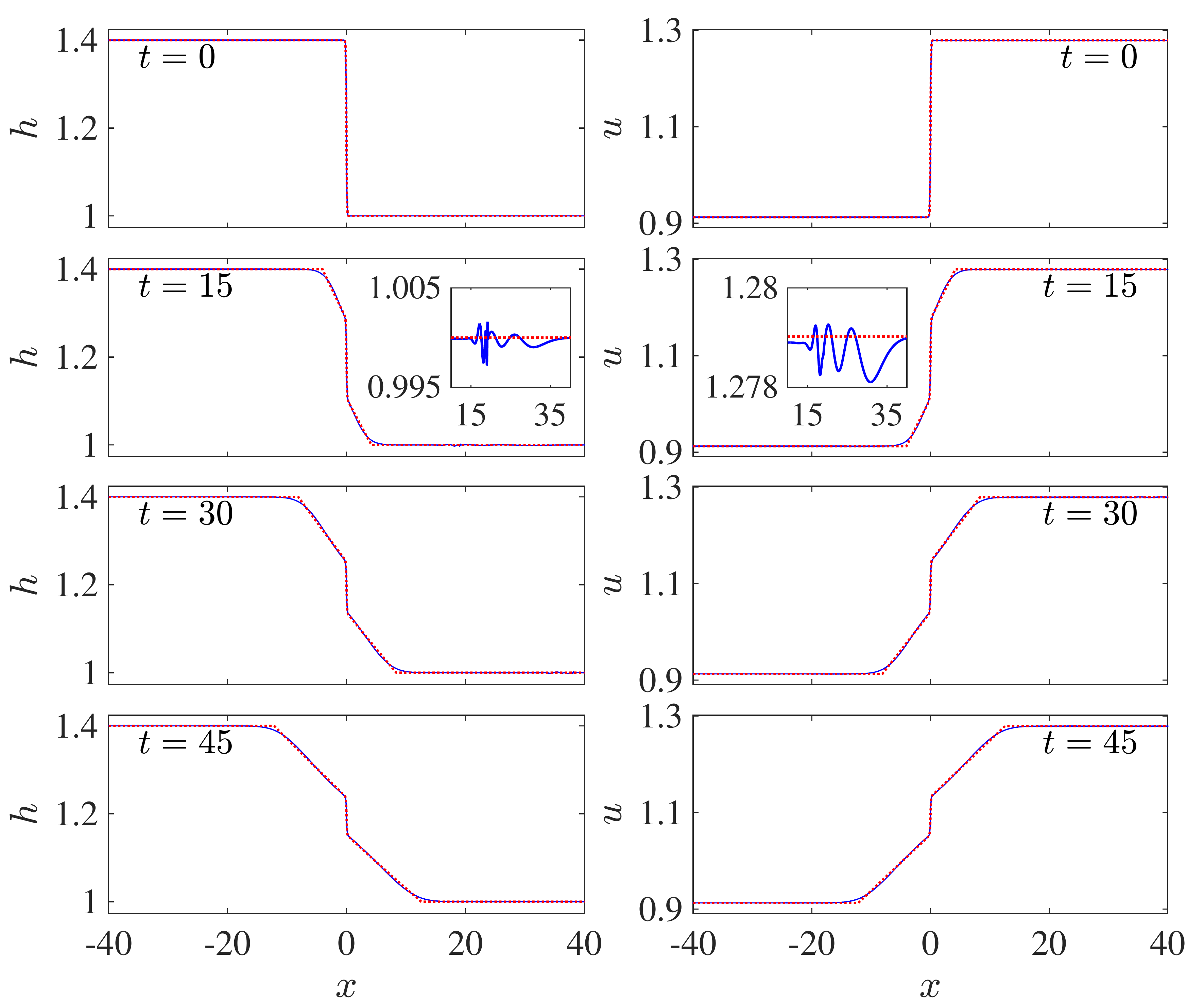}
  \caption{Numerical simulation of the Boussinesq equation for
    expansion shock initial data in physical variables $h$, $u$
    (solid) compared with the the uniform asymptotic approximation
    (dotted) for $H = 1.4$, $\epsilon \approx 0.366$, $\delta =
    0.1$.}
  \label{fig:numerics_h_u}
\end{figure}

\begin{figure}
  \centering
  \includegraphics[scale=0.33333]{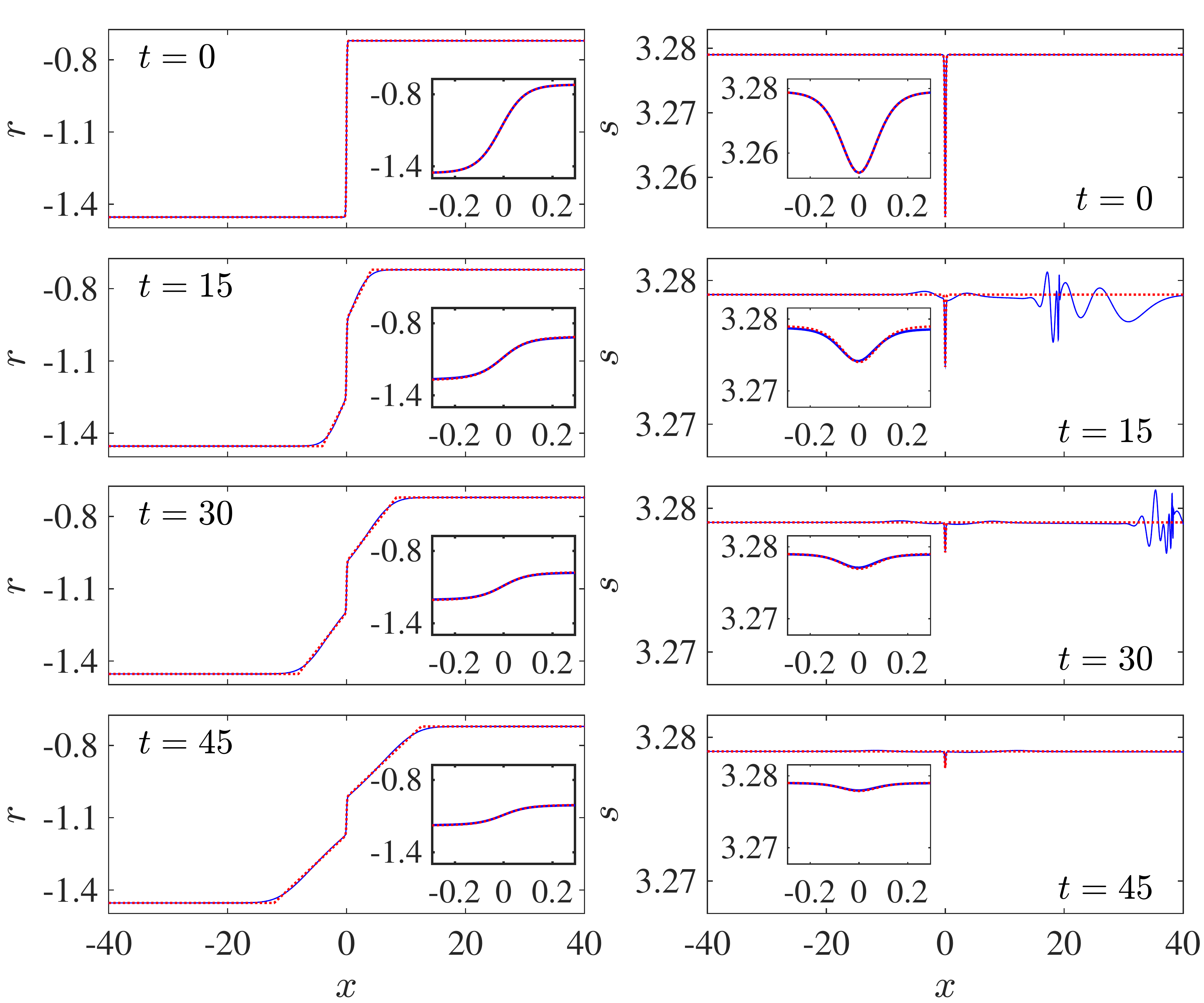}
  \caption{Numerical simulation of the Boussinesq equation for
    expansion shock initial data in the transformed variables $r$, $s$
    (solid) compared with the uniform asymptotic approximation
    (dotted) for $H = 1.4$, $\epsilon \approx 0.366$, $\delta =
    0.1$.  The insets reveal the zoomed in expansion shock structure.
    Note the change in the $s$ amplitude scale for $t > 0$.}
  \label{fig:numerics_r_s}
\end{figure}

It is revealing to examine the evolution of the scaled Riemann
variables $r$ and $s$ in Fig.~\ref{fig:numerics_r_s}.  The variable
$r$ evolves much like $h$ and $u$, with order one changes in
amplitude.  The smooth, decaying expansion shock structure is
accurately resolved by the asymptotic approximation, as shown in the
insets.  The evolution of $s$, on the other hand, is at a much smaller
amplitude scale.  Recall that the RH locus \eqref{RH_locus} leads to
an order $\epsilon^3$ jump in $s$ across an expansion shock.  This
variation in $s$ is not captured by our asymptotic approximation
\eqref{eq:22} and is the source of the dispersive wavepacket that
propagates away from the initial transition.  Note that although the
oscillations appear sharp in the figure, they are smoothly and
accurately resolved by the numerical simulation. Presumably, a higher
order correction to the obtained expansion shock approximation
\eqref{eq:22} would reduce the amplitude of this
wavepacket. Nevertheless, the initial, order $\epsilon^2$ amplitude
dip in $s$ at the transition is apparent and accurately captured by
the asymptotic approximation.

\begin{figure}
  \centering
  \includegraphics[scale=0.33333]{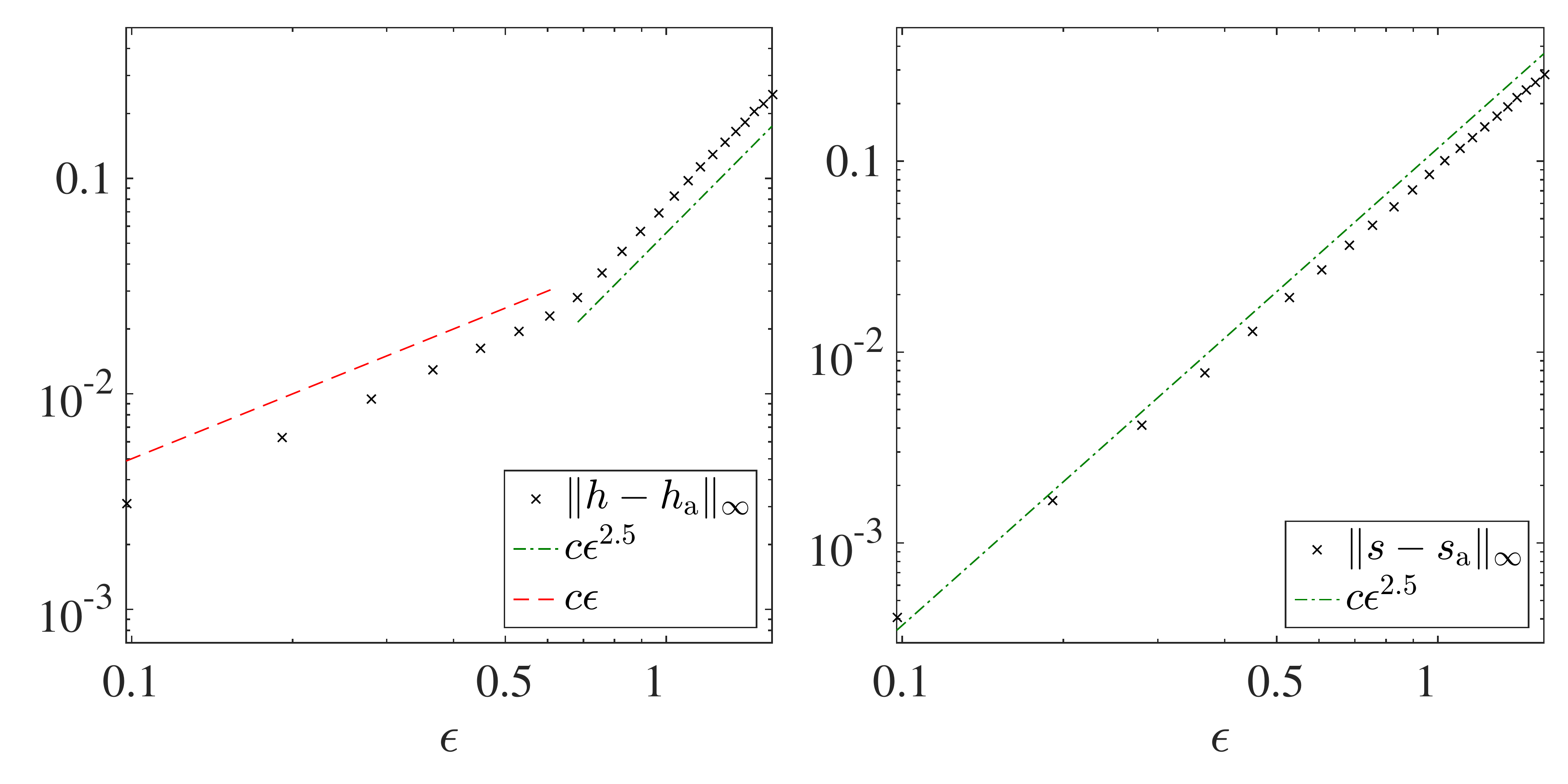}
  \caption{Error analysis of approximate expansion shock solution for
    variable $\epsilon$ and fixed $\delta = 0.1$.  We compare
    simulations of the Boussinesq equations with the asymptotic
    solution \eqref{eq:22} in the infinity norm for $h$ (left) and $s$
    (right).}
  \label{fig:error}
\end{figure}

We undertake an error analysis of the asymptotic expansion shock
solution \eqref{eq:22} by performing numerical simulations with
variable initial jump height parameter $\epsilon$ and fixed transition
width $\delta = 0.1$.  Figure \ref{fig:error} shows a summary of the
results, comparing the infinity norm difference between the asymptotic
approximation, denoted by the subscript ``a'', and the numerical
solution for both $h$ and $s$ as $\epsilon$ is varied.  The errors in
$u$ and $r$ are similar to those for $h$.  The norm difference is
computed across the entire simulation domain, i.e., for $x \in [-L,L]$
and $t \in [0,T]$.  For these simulations $L = 120$, $T = 45$.  Both
$h$ and $s$ show an approximately $\mathcal{O}(\epsilon^{5/2})$
dependence of the error over a portion or all of the jump heights
considered.  The dominant contribution to these errors is due to the
dispersive wavepacket that is generated by the discrepancy in the
approximate initial data (recall the insets in
Fig.~\ref{fig:numerics_h_u}). The error is consistent with the formal
second order accuracy of the asymptotic solution.  It is striking that
the asymptotic solution exhibits small error, even for values of
$\epsilon$ above one. Below $\epsilon = 0.6$, the error in $h$ decays
at a slower rate approximately proportional to $\epsilon$.  This is
because the dominant error contribution now comes from the region
where the non-centered rarefaction waves are matched to the constant
background, eq.~\eqref{eq:59}.  The higher order approximation fails
to resolve this region, which is smoothed in the numerical solution by
dispersion.  This discrepancy is visible in
Fig.~\ref{fig:numerics_h_u}.

\section{Discussion}
\label{sec:discussion}
Decaying expansion shocks were recently identified as robust solutions
to conservation laws of non-evolutionary type that naturally arise in
shallow water theory.  In \cite{el_hoefer_shearer_2016} we studied
these shocks in the framework of the unidirectional BBM equation,
using matched asymptotic expansions.  In the present paper, the
analysis of \cite{el_hoefer_shearer_2016} is extended to a
bi-directional regularized Boussinesq system \cite{whitham_book}.  The
extension to the bi-directional case is more complicated, and has
revealed further structure of expansion shocks exhibiting subtle but
essential features that appear in the second order corrections of the
matched asymptotic expansion, while the first order solution is
equivalent to the BBM expansion shock. The key feature of our analysis
of the Boussinesq expansion shocks is the use of the Riemann
invariants of the underlying ideal shallow water equations as field
variables in the full dispersive system. Another important feature is
the requirement of the balance between two small parameters: the width
$\delta$ of the smoothed Riemann data satisfying the stationary
expansion shock Rankine-Hugoniot conditions and the value of the
initial jump of the water height, measured by $\epsilon$.  The product
$\mu=\epsilon \delta$ then sets the inverse time scale for the
algebraic decay of the expansion shock.

A natural extension of this work is the consideration of jump initial
data that does not lie on the RH locus \eqref{RH_locus}.  An analogous
problem was numerically studied in the context of the BBM equation
\eqref{BBM} corresponding to asymmetric, positive jump initial data
passing through $v = 0$ \cite{el_hoefer_shearer_2016}.  There, an
expansion shock forms accompanied by a sequence of solitary waves, the
number depending upon the asymmetry of the data.  The bi-directional
nature of the Boussinesq equations \eqref{boussinesq1} suggests a
richer set of outcomes.

The comparison of the obtained second order asymptotic formula with
accurate numerical solution of the smoothed Riemann problem for the
Boussinesq system reveals remarkable agreement, even for relatively
large initial jumps, beyond the formal applicability of our asymptotic
analysis.  In conclusion, we note that, in considering more general
initial data, the use of the ``dispersionless'' Riemann invariants as
dependent variables in the full system may give insight into the
structure of solutions, since the interaction between the two fields
occurs primarily through the dispersive terms, except where waves
collide.

% Acknowledgements
\section*{Acknowledgments}

The research of MS and MH is supported by National Science Foundation
grants DMS-1517291 and CAREER DMS-1255422, respectively.

\begin{appendix}
\section*{Appendix}
\label{sec:appendix}

A pseudospectral Fourier spatial discretization with standard fourth
order Runge-Kutta timestepping is utilized.  We discretize the domain
$[-L,L]$ according to $x_n = -L + 2Ln/N$, $n=0,\cdots,N-1$.  In order
to accommodate non-periodic boundary conditions in $h$ and $u$, the
spatial derivatives $g = h_x$, $v = u_x$ are numerically evolved
according to
\begin{equation}
  \label{eq:24}
  \begin{split}
    g_t + (ug)_x + (h v)_x &= 0, \\
    v_t + (uv)_x + g_x - \frac{1}{3} v_{xxt} &=0 .
  \end{split}
\end{equation}
By choosing a sufficiently large domain $L$, the boundary quantities
$|h(\pm L,t) - h_\pm|$ and $|u(\pm L,t) - u_\pm|$ are maintained to
within $10^{-9}$ for the duration of the simulation, therefore $g$ and
$v$ can be treated as localized, periodic functions.  Each
un-differentiated term in eq.~\eqref{eq:24} is spatially localized,
therefore we can compute their derivatives in spectral space, e.g.,
\begin{equation}
  \label{eq:28}
  \mathcal{F}\{(ug)_x\}_n = ik_n \mathcal{F}\{ug\}_n , \quad n = -N/2,
  \cdots, N/2-1,
\end{equation}
where $\mathcal{F}$ is the discrete, finite Fourier series operator,
efficiently implemented via the FFT, and $k_n = n\pi/L$ are the
discrete wavenumbers.  The function $h$ is approximated by an
accumulation of its derivative $g$ according to
\begin{equation}
  \label{eq:44}
  h(x_n,t) = h_- + \mathcal{F}^{-1} \left \{ \tilde{g}(t)
  \right \}_n + \frac{1}{2L}(h_+-h_-) (x_n+L),
\end{equation}
where
\begin{equation}
  \label{eq:51}
  \tilde{g}_n(t) =
  \begin{cases}
    \displaystyle -\sum_{m=-N/2}^{N/2-1} x_m g(x_m,t) & n = 0 \\[6mm]
    \displaystyle \frac{\hat{g}_n(t)}{ik_n} & n \ne 0
  \end{cases} .
\end{equation}
The sum for $n=0$ in \eqref{eq:51} is a trapezoidal approximation of
the integral $\int_{-L}^L x g(x,t) dx$ so that an accurate, efficient
reconstruction of $h$ from $g$ is achieved.  A similar computation is
performed to obtain $u$.

Time evolution is performed on the spectral, Fourier coefficients
using the standard fourth order Runge-Kutta method.  The nonlocal
character of the dispersive term $v_{xxt}/3$ in eq.~\eqref{eq:24} is
not stiff so we use a timestep of $0.002$ and evolve to $t = 45$.  The
domain size is $L = 120$ (Figures \ref{fig:numerics_h_u} and
\ref{fig:numerics_r_s} show only a portion of the domain) and the
Fourier truncation is $N = 2^{14}$.  The accuracy of the numerical
computation is monitored by ensuring that the conserved quantities
$|\int_{-L}^L g(x,t) dx - h_+ + h_-|$, $|\int_{-L}^L v(x,t) dx - u_+ +
u_-|$ are maintained to less than $10^{-14}$ and the Fourier
components $|\mathcal{F}\{g\}_n|$, $|\mathcal{F}\{v\}_n|$, $n =
-N/2,\ldots,N/2-1$ decay to about $10^{-8}$, within the expected value
given boundary deviations of about $10^{-9}$.

\end{appendix}

% References


\begin{thebibliography}{00}

\bibitem{el_hoefer_shearer_2016}  \textsc{G.~A.~El, M.A. Hoefer} and \textsc{M.~Shearer},
Expansion shock waves in regularized shallow water theory. \emph{Proc. Roy. Soc. London}  472: 20160141 (2016)

\bibitem{whitham_book} \textsc{G.~B.~Whitham},
\emph{ Linear and Nonlinear Waves.} (Wiley, New York,  1974).

\bibitem{bona_etal_2002} 
\textsc{J.~L.~ Bona, M.~Chen} and \textsc{J.~C.~Saut},
 Boussinesq equations and other systems for small--amplitude long
  waves in nonlinear dispersive media. {I}: derivation and linear theory.
\emph{J Nonlinear Sci.} 12: 283-318 (2002).

\bibitem{el_hoefer_dsw_2016} \textsc{G.~A.~El} and \textsc{M.A. Hoefer}, Dispersive shock waves and modulation theory, \emph{Physica D}  33: 11-65 (2016).

\bibitem{el_hoefer_shearer_mkdv_2016}   \textsc{G.~A.~El, M.A. Hoefer} and \textsc{M.~Shearer}, Dispersive and diffusive-dispersive shock waves for non-convex conservation laws, \emph{SIAM Review} (2017), accepted

\bibitem{lax} \textsc{P.~D.~Lax}, Hyperbolic Systems of Conservation Laws II, \emph{Comm. Pure Appl. Math.}
 10: 537-566 (1957).
 
\bibitem{bressan} \textsc{A.~Bressan}, \emph{ Hyperbolic Systems of Conservation Laws: The One-Dimensional Cauchy Problem.} (Oxford Univ. Press, 2000). 

\bibitem{milewski} \textsc{F.~Dias} and \textsc{P.~Milewski}, On the
  fully-nonlinear shallow-water generalized Serre equations,
  \emph{Phys. Lett. A}  374: 1049-1053 (2010).


\end{thebibliography}
\end{document}